# Ferroelectric Solitons Crafted in Epitaxial Bismuth Ferrite Superlattices

V. Govinden[1]†, P. R. Tong[2]†, X. Guo[3,4,5]†, Q. Zhang[1], S. Mantri[6], S. Prokhorenko[6], Y. Nahas[6], Y. Wu[3,5], L. Bellaiche[6], H. Tian[2*], Z. Hong[3,5*], D. Sando[1*] and V. Nagarajan[1*]

†- these authors contributed equally

*- To whom correspondence should be addressed

E-mail: hetian@zju.edu.cn, hongzijian100@zju.edu.cn, daniel.sando@unsw.edu.au, nagarajan@unsw.edu.au

[1]School of Materials Science and Engineering, University of New South Wales, Sydney, NSW 2052, Australia

[2]Center of Electron Microscope, State Key Laboratory of Silicon Material, School of Material Science & Engineering, Zhejiang University, Hangzhou 310027, China

[3]School of Materials Science and Engineering, Zhejiang University, Hangzhou, Zhejiang 310027, China

[4]Institute of Advanced Semiconductors & Zhejiang Provincial Key Laboratory of Power Semiconductor Materials and Devices, Hangzhou Innovation Center, Zhejiang University, Hangzhou, Zhejiang 311200, China

[5]Cyrus Tang Center for Sensor Materials and Applications, State Key Laboratory of Silicon Materials, Zhejiang University, Hangzhou, Zhejiang 310027, China

[6]Physics Department and Institute for Nanoscience and Engineering, University of Arkansas, Fayetteville AR 72701, USA

**Abstract**

In ferroelectrics, complex interactions among various degrees of freedom enable the condensation of topologically protected polarization textures. Known as ferroelectric solitons, these particle-like structures represent a new class of materials with promise for beyond-CMOS technologies due to their ultrafine size and sensitivity to external stimuli. Such polarization textures have scarcely been reported in multiferroics. Here, we report a range of soliton topologies in $(BiFeO_3)_m/(SrTiO_3)_m$ superlattices. High-resolution piezoresponse force microscopy and Cs-corrected high-angle annular dark-field scanning transmission electron microscopy reveal a zoo of topologies, and polarization displacement mapping of planar specimens reveals center-convergent/divergent topological defects as small as 3 nm. Phase field simulations verify that some of these topologies can be classed as "bimerons", with a topological charge of ±1, and first-principles-based effective Hamiltonian computations show that the co-existence of such structures can lead to non-integer topological charges, a first observation in a $BiFeO_3$-based system. Our results open new opportunities in multiferroic topotronics.

Ferroelectrics are known to exhibit strong coupling of strain to electric polarization. This phenomenon enables the formation of exotic domain structures which can display emergent properties useful for information storage and sensing devices. Modern developments in fabrication techniques have enabled flexible tailoring of strain, chemical and electrical boundary conditions to achieve a new paradigm of nanoscale complex polarization textures, including topologically protected nontrivial states[1-4]. When such topologically protected polarization textures condense to ultrafine size dispersed in a parent medium, they can be considered as particle-like objects known as ferroelectric solitons. An example of such three-dimensional ferroelectric solitons are spherical domains (and their transitional states), which possess homogeneously polarized cores surrounded by a curling polarization forming a curved outer shell. The strong polarization and strain gradients which exist at the unit cell level within these structures results in extremely high local crystalline anisotropy and even symmetries that are energetically unfavorable in the parent bulk. This has a significant implication: the polarization curling dramatically raises the internal energy, with two consequences: i) the size of such topological objects is restricted to the nanoscale (i.e., 1-10 nm), and ii) these objects are highly sensitive to external stimulus. These virtues make ferroelectric solitons prime candidates for low-energy and high-density nanoelectronics[5].

Ferroelectric solitons have various forms ranging from electrical bubbles[6-8], to polar bubble skyrmions[9, 10] and more recently – thus far only theoretically predicted – hopfions[11]. Other transitional topologies such as merons[12], bimerons[13] and disclinations[13, 14] have also been observed. Although spherical ferroelectric topologies were theoretically predicted more than two decades ago,[15, 16] and flux closure domain structures studied previously[17-19], the field has arguably not seen such a dramatic surge in efforts[20-22] since the demonstration of stable polarization vortex arrays in epitaxial lead titanate-strontium titanate superlattices by Yadav[23] and Tang[1]. Following these reports, it was shown that ferroelectric/dielectric superlattices could be tuned to fabricate skyrmion arrays[10] with emergent chiral[24], local negative permittivity[25] and conduction properties[26]. However, such topologies have also been found in simple ferroelectric sandwich heterostructures[27], revealing that ultimately the delicate interplay between the electrical and mechanical boundary conditions drives their formation[16, 28, 29].

Whilst solitons with complex non-trivial topologies have been demonstrated in pure ferroelectrics[7, 10] and type-II multiferroics[30] (where the polarization is the secondary order parameter), they have remained elusive in type-I multiferroics. These latter materials also harbor coexisting ferroelectric and magnetic orders, but here the ferroelectric polarization is the primary order parameter meaning that the local crystalline anisotropy (i.e., strain) energy plays a more dominant role than is the case for their type-II cousins. A major type-I room temperature multiferroic candidate is bismuth ferrite ($BiFeO_3$ – BFO) which boasts a plethora of appealing properties including multiferroic[31, 32], photovoltaic[33-35], piezoelectric[36], domain-wall conduction[37, 38] and optoelectronic responses[39]. In this context, the observation of spherical and transitional topologies such as skyrmions, polar vortex arrays, merons and bubble domains in BFO would undoubtedly have wide-reaching implications, both fundamentally and practically[40]. Although epitaxial BFO heterostructures have been engineered to demonstrate writable vortex cores[41], center convergent and quad-domain structures[42, 43] or self-assembled flux closure arrays[44], the observation of topological solitons is still to be achieved. A natural question thus arises: how can one craft the hitherto evasive solitons in BFO?

To address this, one must consider the requirements for the creation of such a polarization configuration in a polar material[45]. The challenge is to engineer a ferroelectric on the brink, where continuous polarization rotation is achieved without pushing the system to form symmetry breaking Ising domain walls[46].

Here, we report the deterministic stabilization of the complex topological phases in epitaxial $BiFeO_3$ (BFO)–$SrTiO_3$ (STO) superlattices fabricated on (001)-oriented $LaAlO_3$ (LAO) substrates using pulsed laser deposition. The superlattices display sharp interfaces as revealed by Cs-corrected high-angle annular dark-field scanning transmission electron microscopy (HAADF-STEM) and atomic resolution energy dispersive spectroscopy (EDS) mapping. Piezoresponse force microscopy (PFM) reveals a diverse range of non-trivial spherical domains, hinting towards the coexistence of a zoo of ferroelectric solitons. Atomic-scale polarization displacement mapping of planar HAADF-STEM specimens was thus carried out. These measurements confirmed the existence of a range of topological structures, from bubbles to bimerons to possibly polar skyrmions. These latter defects show both center-convergent and center-divergent polarization profiles with sizes as small as 3 nm. The experimental results are verified with first-principles-based effective Hamiltonian predictions and phase-field simulations, which find that the origin for the formation of these non-trivial topologies is the competing electrical and mechanical boundary conditions. Critically, the simulations find that the co-existing topological phases can each individually possess either fractional or integer topological charge. The discovery of such ultrafine topologically protected states in multiferroic BFO unlocks an uncharted platform with new degrees of freedom, i.e., control by both electric and magnetic fields.

Figure 1A graphically depicts the basis of our approach. In ferroelectric/dielectric/ferroelectric structures or ferroelectric/dielectric superlattices, the incorporation of a dielectric layer (often STO) leads to the accumulation of bound charges at the interfaces thus increasing the system's free energy. The out-of-plane polarization is thus forced to curl in an in-plane direction to reduce the energy and can thus stabilize ferroelectric solitons. In PbTiO$_3$ (PTO)-based systems, to achieve polarization curling, the aim is to force the naturally oriented out-of-plane pointing polarization to tilt towards the film plane. In BFO, on the other hand, we use the opposite principle. This is because bulk BFO crystallizes in rhombohedral symmetry with polarization along [111] direction. We thus need to impose a compressive strain to push the polarization vector towards the surface normal. For this reason, we use LAO as our substrate, enabling us to impose an in-plane compressive stress that stabilizes an out-of-plane polarization towards the [001] direction (i.e., the so-called T-phase). This is counterbalanced by a depolarization field created by the insertion of a STO spacer. The spacer has two key effects- (i) it breaks polarization and structural continuity and (ii) it provides the mechanical compatibility at the interface to allow the polarization to curl[47].

Although the principle seems simple, growing ultra-smooth layers of epitaxially strained BFO to several tens of nm is not trivial. This is where our unique pulsed laser deposition (PLD) chamber system – with a large substrate to target separation (~10 cm) – comes into play. First, this large distance results in low incident flux, enabling controlled ultra-slow layer-by-layer growth of BFO under compressive strain (additional data available on request). Second, the low flux at high temperatures can achieve self-regulated growth of tetragonal like (T-like) BFO to thicknesses up to 60 nm with no mixed phase formation[48]. This ability to fabricate superlattices wherein the "long-range" in-plane compressive strain can be sustained (i.e., canting the polarization towards the [001] direction of BFO) – somewhat surprisingly as we will see later – is a key first step towards the realization of curling polarization textures in this material.

The next step is to identify the optimal superlattice design. Even a slight deviation from the optimized thickness induces relaxation mechanisms[49]. The thickness of the BFO layer is pivotal: each layer must be thin enough to maintain the imposed "macroscopic" strain and dipolar coupling at the interface[50], and couple across the STO spacers, but not to be locally influenced by intrinsic size effects (Fig. 1A). By carrying out a detailed study on the influence of individual layer thickness (additional data available on request) we found the optimal configuration for creating topological textures to be (BFO$_7$/STO$_4$)$_{10}$.

Structural and chemical characterization of the above system is summarized in Fig. 1A. A representative cross-sectional STEM image within the heterostructure $(BFO_7/STO_4)_{10}$ and its corresponding atomic resolution EDS map (see below) show atomically and chemically sharp heterointerfaces with no interdiffusion (Comprehensive HAADF- STEM and EDS analysis of the complete structure can be provided on request)). Since HAADF is sensitive to the atomic number, the individual BFO and STO layers are seen as alternate bright and dark contrast with a thickness of 7 and 4 unit-cells (u.c.), respectively, in agreement with the intended superlattice design. The well-aligned atomic columns demonstrate the high-quality epitaxial configuration with the absence of any defects. The EDS maps reveal distinct Ti (cyan), Fe (green), Sr (yellow) and Bi (red) atomic positions with no sign of interdiffusion confirming perfect coherent stacking of the BFO and STO layers.

Figure 1B presents symmetrical XRD reciprocal space mapping (RSM) of a $(BFO_7/STO_4)_{10}$ superlattice near the 002 reflection of film and substate, revealing various peaks along the out-of-plane direction ($Q_z$). In addition to the 002 LAO substrate peak, the next brightest spot corresponds to the out-of-plane periodicity of the superlattice with an average out-of-plane parameter of $4.00 \pm 0.01$ Å. Simulations of the diffraction pattern using a custom-made MATLAB program[51] show that the data are consistent with out-of-plane lattice parameters of BFO (STO) of 4.07 Å (3.91 Å) (additional data available on request). The sharp interfaces between the BFO and STO layers lead to additional superlattice reflections such as the $SL_{-1}$, corresponding to the 11 u.c. repeat length. A pertinent feature of this dataset is the breadth in the horizontal direction of the main film peak, which could arise from increased local mosaicity, defects, or strain gradients. Since we do not observe chemical defects or dislocations from STEM, we rule out the first two possible effects, implying that significant strain gradients exist within the superlattice structure. The origins of these strain gradients will become clear in Fig. 2. Finally, a peak with narrow horizontal width is detected at lower $Q_z$ values which is indexed as tetragonal-like T-BFO, likely stabilized in the layers closer to the substrate. The average in-plane lattice parameter of the $(BFO_7/STO_4)_{10}$ is $3.92 \pm 0.02$ Å, close to the value of bulk STO (additional data available on request). Thus, the c/a ratio of the entire stack is ~1.04 implying that the BFO is not T-like, but more strained R-like due to some degree of strain relaxation. We return to this key observation later when we discuss first-principles-based computations. Local measurements of the lattice parameters using STEM are discussed later.

The critical role played by the LAO substrate now becomes apparent. First, the imposed in-plane compressive strain favors the BFO layer's polarization to be aligned predominantly out of plane. But at the same time, the STO spacer interrupts the polar continuity, inducing strong polarization gradients near the interfaces to minimize the overall increased electrostatic energy costs. These combined effects drive BFO into a state of absolute polar frustration: it can have neither long range out-of-plane polarization, which is prohibited by the cost of depolarization, nor full in-plane tilting, this being restricted by the underlying substrate.

Next, the ferroelectric domain structure was imaged using high-resolution dual amplitude resonance tracking (DART) PFM. The amplitude and phase PFM images (Fig. 1C) demonstrate intricate nanoscale domain configurations. Interestingly, we observe topological features with remarkable semblance to those found by Milde *et al.* for magnetic skyrmions in $Fe_{1-x}Co_xSi$ (Ref. 52). A magnified scan (Fig. 1D) reveals the occurrence of sub-20 nm nanoscale domains, labelled (1) and outlined by white dashed rectangles. These nanodomains show blurry amplitude contrast and a faint upward domain phase of the domain wall. Previously, such features were ascribed to either bubble domains[7, 53] or skyrmions[8, 10]. Furthermore, transitional topological defects such as bimerons[13] and disclinations[13] (depicted as (2) and (3) respectively in Figs. 1, D and E, are also identified. Since the resolution of PFM limits our ability to image any topological feature smaller than ~20 nm, this technique alone cannot discern bubbles or skyrmions. The main feature distinguishing polar skyrmions from bubbles is an additional polarization vortex along the circumferential axis, which cannot be imaged in PFM. Hence, we next characterize these 3-dimensional structures at the atomic scale using Cs-STEM.

Figure 2A show the planar-sectional Cs-STEM of epitaxial $(BFO_7/STO_4)_{10}$ superlattices which reveal arrays of circular feature with typical diameters of ~3.5 nm, whose size distribution is shown in Fig. 2B. The application of a displacement vector-mapping algorithm on both the cross-sectional and planar view HAADF-STEM images provides direct visualization of atomic-scale polarization displacement within the superlattices (Figs. 2, D and E) (additional data available on request). The planar view HAADF-STEM image and corresponding vector displacement mapping demonstrated a variety of topological states. In Figs. 2D, Ⅰ and Ⅱ, we confirm a coexistence of in-plane center-divergent and center-convergent polar textures, which take the form of circular domains in low-magnification STEM images. The bimeron with both center-divergent and anti-vortex polar structures is identified (Fig. 2D, Ⅲ), which appears as a fusion of two bubble domains in STEM image. Polar regions with anti-parallel (up *vs.* down) polarizations are found in the cross-sectional HAADF-STEM vector displacement mapping (Fig. 2E, Ⅰ), with polarization curling occurring near the BFO/STO interfaces. In Fig. 2E, Ⅱ, we reveal a meron-type region with a core consisting of out-of-plane polarization, and as one moves away from the center, the polarization gradually changes into in-plane directions. The combination of these out-of-plane and in-plane polar configurations shows that we have three-dimensional polar solitons in our $(BFO_7/STO_4)_{10}$ superlattices.

While this observation is qualitatively similar to the skyrmions observed in $(PTO_{16}/STO_{16})_8$ superlattices[10], our system shows two important distinctions. First, the objects in our $(BFO_7/STO_4)_{10}$ are *not* completely confined within the BFO layers: their polar order also partially exists in the nominally paraelectric STO spacer. We attribute this effect to the strong electrostatic coupling between the thin paraelectric and ferroelectric layers which causes dipole rotation in the confined STO[54]. Second, the smaller thickness of BFO ferroelectric layer (7 u.c.) leads to significantly smaller characteristic size of the topological objects (~3.5 nm) (Fig. 2B) as compared to the size observed in PTO (16 u.c.)/STO superlattices (8 nm)[10]. Figure 2C shows the statistics of *c/a* ratios of BFO and STO. The lattice parameters of unit cells were calculated by fitting each atom site by a spherical Gaussian using an algorithm in MATLAB. The average *c/a* ratio (1.0540) implies that the BFO is not T-like, fits with but is slightly larger than the value of the previously reported *c/a* ratio of R-phase BFO[55].

The discrepancy between the size and type of topological features observed by STEM vis-à-vis PFM could be attributed to various factors: (i) We have identified a whole range of different topological structures, meaning that one should not be surprised that different techniques see different topological objects. (ii) In PFM, the trailing field effect of the tip[18] where even the slight pressure and electric field enlarge the topological state, and/or (iii) removal/thinning of the substrate for making planar view STEM sample, thereby dramatically affecting the boundary conditions. Arguably, case (iii) is less likely, as our previous work on bubbles does not find a difference in the bubble size for free-standing *vs.* constrained samples[56]. Moreover, no difference was observed between PFM and STEM for skyrmions in the PTO system. We have, on the other hand, shown that bubbles are extremely sensitive to applied scanning probe microscopy (SPM) pressure and scanning field[7, 53].

To gain further insight into the topological polarization configuration and formation mechanism of the polar states observed experimentally in our superlattices, phase-field simulations were performed (additional data available on request). Figure 3 shows the typical polar structures and topological features of $(BFO_7/STO_4)_8$ superlattices. As seen in Fig. 3A, the in-plane polarizations show donut-like circular patterns, similar to the polar skyrmion structure observed in the PTO/STO system[10]. Meanwhile, the detailed polar mapping demonstrates that the observed structure in this system is distinctly different from the polar skyrmions in the PTO/STO superlattice. As shown in Fig. 3B, the in-plane polarization forms a combination of center convergent dipole configuration with anti-vortex like structure, which is the characteristic of a "bimeron" structure, consistent with the experimental observation by STEM (Fig. 2D). The topological feature is characterized by calculating the Pontryagin density, $P_d = \vec{P} \cdot (\frac{\partial \vec{P}}{\partial x} \times \frac{\partial \vec{P}}{\partial y})$. It can be observed that the bubble-like structure shows a circular distribution of the Pontryagin density (Fig. 3C). The magnified view of the Pontryagin density distribution for one bimeron demonstrates a circular structure (Fig. 3D), while the surface integration of the circle gives a topological structure of -1, confirming that such structures are more like bimerons. Furthermore, both center divergent and center convergent type structures are observed in this system (additional data on request), in good agreement with the experimental observation. The cross-section view is further plotted, as given in Figs. 3, E and F. A 180º domain wall like structure is observed, with alternating positive and negative out-of-plane polarization (Fig. 3E). The polarization vector is plotted, as demonstrated in Fig. 3F, showing the formation of polar vortex like structure. While it is interesting to note that in the BFO layer, both *R*-like and *T*-like regions are identified, in agreement with the experimental observations.

Further computational approaches also give strong support for the formation of topological objects under moderate compressive strain in our heterostructures, as explained next. A first-principles-based effective Hamiltonian[57, 58] was used to study thin film BFO. A 16 u.c. thick (001) oriented BFO film with an initial 109° domain structure was placed under mechanical boundary condition of epitaxial substrate strain in the compressive range varying up to -5% with an open-circuit-like electrical boundary condition. Multidomain structures are known to be preferred over monodomains under open circuit like electrical boundary conditions[16]. The specific multidomain structure of 109° domain type was chosen as it has been the most experimentally observed domain wall in bulk $BiFeO_3$ and has the lowest domain wall energy as predicted by first-principles-based calculations[59-61]. The film is modelled by a supercell periodic in the *x* and *y* directions. Monte Carlo simulations were run for 50,000 steps at 10 K for each substrate strain. It was observed that the initial 109° domains changed into a topologically non-trivial structure (Fig. 4A), that had a vortex in the *x-z* plane with convergent/divergent domain walls in the *x-y* plane located at the circumference of the vortices.

These features seen in the computed polar structure agree with HAADF-STEM experiments. Namely, effective Hamiltonian simulations reproduce the zig-zag vortex pattern observed in cross-sectional HAADF-STEM images (pattern I in Fig. 2E) while convergent/divergent lines at the BFO interfaces in Fig. 4A support the formation of center convergent/divergent points observed in Fig. 2D. The vortex occurred in a moderate compressive strain region of [-3%, -2%] beyond which it is destroyed. The c/a value for the substrate strains that resulted in a vortex lies between 1.077 to 1.050 at the misfit strains of -3% and -2%, respectively. To analyze the topology of the resultant dipolar structure, we have calculated the distribution of the Pontryagin's charge density ($\rho_{Sk} = \boldsymbol{n}.(\partial_x \boldsymbol{n} \times \partial_y \boldsymbol{n})$, where $\boldsymbol{n}$ denotes normalized electric dipoles) in the *x-z* plane. The distribution of the resulting density (colored plaquettes) and the normalized dipoles (arrows) are shown in Fig. 4B where red and blue circles indicate examples of meron and antimeron lines. These lines are found to carry a *fractional* Skyrmion number. Particularly, the meron indicated by a red circle in Fig. 4B has a Skyrmion number of 0.54 while the anti-merons outlined in blue carry charges of -0.43 and -0.5. Notably, despite the presence of both positively and negatively charged merons, the integral Skyrmion number ($\int (dx\, dy\, \rho_{Sk})$) over the simulation supercell was found to be equal to one.

## Conclusion

In summary, we have crafted polar soliton structures in epitaxial multiferroic BFO–STO superlattices. This demonstration of these previously elusive topological states in this material system is anticipated to have far reaching implications on the topological landscape of topological polar/spin textures in multiferroics. Given that the various Dzyaloshinskii-Moriya interactions – which govern the weak ferromagnetic moment and the spin cycloid in BFO (Refs. 58, 62, 63) – are driven by strain and local symmetry, one can anticipate that solitons in BFO could reveal enhanced local ferromagnetic moments, depressed magnetic transition temperatures, and/or enhanced electrical conductivity. Further, similar to how domain walls in materials such as TbMnO$_3$ can harbor exotic electronic and magnetic states[64], our solitons may indeed constitute a fundamentally different multiferroic phase of BFO. These findings are clearly just the tip of the iceberg; we hope that our results will motivate practitioners and theorists in the field to dig deeper into these superlattice systems. Future work will require elucidating the specific role of the local symmetry changes within the solitons and how it influences the local polarization dynamics, weak ferromagnetic moment, optical behavior, and transport responses- all functionalities that can be used in next generation nanoscale devices.

## Acknowledgments

The research at University of New South Wales (UNSW) is supported by DARPA Grant No. HR0011727183-D18AP00010 (TEE Program), partially supported by the Australian Research Council Centre of Excellence in Future Low-Energy Electronics Technologies (project number CE170100039) and funded by the Australian Government. This work is also supported by the National Natural Science Foundation of China (grant No. 92166104 & 12125407), the Zhejiang Provincial Natural Science Foundation (LD21E020002), the Joint Funds of the National Natural Science Foundation of China (U21A2067), the National Key Research and Development Program of China (No. 2021YFA1500800). ZH also gratefully acknowledge a start-up grant from Zhejiang University. Q. Z. acknowledges the support of a Women in FLEET Fellowship. The research at University of Arkansas is also supported by the Vannevar Bush Faculty Fellowship (VBFF) Grant No. N00014-20-1-2834 from the Department of Defense and Arkansas Research Alliance, ARO Grant No. W911NF-21-1-0113 and ARO grant number W911NF-21-2-0162 (MURI-ETHOS). We also acknowledge the Arkansas High Performance Computing Center (AHPCC).

## Bibliography

1.      Y. L. Tang, Y. L. Zhu, X. L. Ma, A. Y. Borisevich, A. N. Morozovska, E. A. Eliseev, W. Y. Wang, Y. J. Wang, Y. B Xu, Z. D. Zhang, S. J. Pennycook, *Science* **348**, 547-551 (2015).

2.      C.-L. Jia, K. W. Urban, M. Alexe, D. Hesse, I. Vrejoiu, *Science* **331**, 1420-1423 (2011).

3.      S. Chen, S. Yuan, Z. Hou, Y. Tang, J. Zhang, T. Wang, K. Li, W. Zhao, X. Liu, L. Chen, L. W. Martin, Z. Chen, *Adv. Mater.* **33**, 2000857 (2021).

4.      X. Guo, L. Zhou, B. Roul, Y. Wu, Y. Huang, S. Das, Z. Hong, *Small Methods* 2200486 (2022).

5.      G. Tian, W. D. Yang, X. S. Gao, J.-M. Liu, *APL Materials* **9**, 020907 (2021).

6. B.-K. Lai, I. Ponomareva, I. I. Naumov, I. Kornev, H. Fu, L. Bellaiche, G. J. Salamo, *Phys. Rev. Lett.* **96**, 137602 (2006).

7. Q. Zhang, L. Xie, G. Liu, S. Prokhorenko, Y. Nahas, X. Pan, L. Bellaiche, A. Gruverman, N. Valanoor, *Adv. Mater.* **29**, 1702375 (2017).

8. J. Yin, H. Zong, H. Tao, X. Tao, H. Wu, Y. Zhang, L.-D. Zhao, X. Ding, J. Sun, J. Zhu, J. Wu, S. J. Pennycook, *Nat. Commun.* **12**, 3632 (2021).

9. Y. Nahas, S. Prokhorenko, L. Louis, Z. Gui, I. Kornev, L. Bellaiche, *Nat. Commun.* **6**, 8542 (2015).

10. S. Das, Y. L. Tang, Z. Hong, M. A. P. Gonçalves, M. R. McCarter, C. Klewe, K. X. Nguyen, F. Gómez-Ortiz, P. Shafer, E. Arenholz, V. A. Stoica, S. L. Hsu, B. Wang, C. Ophus, J. F. Liu, C. T. Nelson, S. Saremi, B. Prasad, A. B. Mei, D. G. Schlom, J. Íñiguez, P. García-Fernández, D. A. Muller, L.-Q. Chen, J. Junquera, L. W. Martin, R. Ramesh, *Nature* **568**, 368-372 (2019).

11. I. Luk'yanchuk, Y. Tikhonov, A. Razumnaya, V. M. Vinokur, *Nat. Commun.* **11**, 2433 (2020).

12. Y. J. Wang, Y. P. Feng, Y. L. Zhu, Y. L. Tang, L. X. Yang, M. J. Zou, W. R. Geng, M. J. Han, X. W. Guo, B. Wu, X. L. Ma, *Nat. Mater.* **19**, 881-886 (2020).

13. Y. Nahas, S. Prokhorenko, Q. Zhang, V. Govinden, N. Valanoor, L. Bellaiche, *Nat. Commun.* **11**, 5779 (2020).

14. L. Lu, Y. Nahas, M. Liu, H. Du, Z. Jiang, S. Ren, D. Wang, L. Jin, S. Prokhorenko, C.-L. Jia, L. Bellaiche, *Phys. Rev. Lett.* **120**, 177601 (2018).

15. I. I. Naumov, L. Bellaiche, H. Fu, *Nature* **432**, 737 (2004).
16. I. Kornev, H. Fu, L. Bellaiche, *Phys. Rev. Lett.* **93**, 196104 (2004).

17. A. Schilling, D. Byrne, G. Catalan, K. G. Webber, Y. A. Genenko, G. S. Wu, J. F. Scott, J. Gregg, *Nano Lett.* **9**, 3359-3364 (2009).

18. N. Balke, S. Choudhury, S. Jesse, M. Huijben, Y. H. Chu, A. P. Baddorf, L.-Q. Chen, R. Ramesh, S. V. Kalinin, *Nat. Nanotechnol.* **4**, 868-875 (2009).

19. B. J. Rodriguez, X. S. Gao, L. F. Liu, W. Lee, I. I. Naumov, A. M. Bratkovsky, D. Hesse, M. Alexe, *Nano Lett.* **9**, 1127-1131 (2009).

20. X. Li, C. Tan, C. Liu, P. Gao, Y. Sun, P. Chen, M. Li, L. Liao, R. Zhu, J. Wang, Y. Zhao, L. Wang, Z. Xu, K. Liu, X. Zhong, J. Wang, X. Bai, *PNAS USA* **117**, 18954-18961 (2020).

21. C. Tan, Y. Dong, Y. Sun, C. Liu, P. Chen, X. Zhong, R. Zhu, M. Liu, J. Zhang, J. Wang, K. Liu, X. Bai, D. Yu, X. Ouyang, J. Wang, P. Gao, Z. Luo, J. Li, *Nat. Commun.* **12**, 4620 (2021).

22. P. Chen, X. Zhong, J. A. Zorn, M. Li, Y. Sun, A. Y. Abid, C. Ren, Y. Li, X. Li, X. Ma, J. Wang, K. Liu, Z. Xu, C. Tan, L. Chen, P. Gao, X. Bai, *Nat. Commun.* **11**, 1840 (2020)

23. A. K. Yadav, C. T. Nelson, S. L. Hsu, Z. Hong, J. D. Clarkson, C. M. Schlepütz, A. R. Damodaran, P. Shafer, E. Arenholz, L. R. Dedon, D. Chen, A. Vishwanath, A. M. Minor, L.-Q. Chen, J. F. Scott, L. W. Martin, R. Ramesh, *Nature* **530**, 198-201 (2016).

24. P. Shafer, P. García-Fernández, P. Aguado-Puente, A. R. Damodaran, A. K. Yadav, C. T. Nelson, S.-L. Hsu, J. C. Wojdeł, J. Íñiguez, L. W. Martin, E. Arenholz, J. Junquera, R. Ramesh, *PNAS USA* **115**, 915-920 (2018).

25. S. Das, Z. Hong, V. A. Stoica, M. A. P. Gonçalves, Y. T. Shao, E. Parsonnet, E. J. Marksz, S. Saremi, M. R. McCarter, A. Reynoso, C. J. Long, A. M. Hagerstrom, D. Meyers, V. Ravi, B. Prasad, H. Zhou, Z. Zhang, H. Wen, F. Gómez-Ortiz, P. García-Fernández, J. Bokor, J. Íñiguez, J. W. Freeland, N. D. Orloff, J. Junquera, L.-Q. Chen, S. Salahuddin, D. A. Muller, L. W. Martin, R. Ramesh, *Nat. Mater.* **20**, 194-201 (2021).

26. L. Han, C. Addiego, S. Prokhorenko, M. Wang, H. Fu, Y. Nahas, X. Yan, S. Cai, T. Wei, Y. Fang, , H. Liu, D. Ji, W. Guo, Z. Gu, Y. Yang, P. Wang, L. Bellaiche, Y. Chen, D. Wu, Y. Nie, X. Pan, *Nature* **603**, 63-67 (2022).

27. D. Rusu, J. J. P. Peters, T. P. A. Hase, J. A. Gott, G. A. A. Nisbet, J. Strempfer, D. Haskel, S. D. Seddon, R. Beanland, A. M. Sanchez, M. Alexe, *Nature* **602**, 240-244 (2022).

28. C. T. Nelson, B. Winchester, Y. Zhang, S.-J. Kim, A. Melville, C. Adamo, C. M. Folkman, S.-H. Baek, C.-B. Eom, D. G. Schlom, L.-Q. Chen, X. Pan, *Nano Lett.* **11**, 828-834 (2011).

29. I. Ponomareva, I. Naumov, L. Bellaiche, *Phys. Rev. B* **72**, 214118 (2005).

30. S. Seki, X. Z. Yu, S. Ishiwata, Y. Tokura, *Science* **336**, 198-201 (2012).

31. D. Sando, B. Xu, L. Bellaiche, V. Nagarajan, *Appl. Phys. Rev.*, **3**, 011106 (2016).

32. N. A. Spaldin, R. Ramesh, *Nat. Mater.* **18**, 203-212 (2019).

33. T. Choi, S. Lee, Y. J. Choi, V. Kiryukhin, S.-W. Cheong, *Science* **324**, 63-66 (2009).

34. D. Sando, Y. Yang, C. Paillard, B. Dkhil, L. Bellaiche, V. Nagarajan, *Appl. Phys. Rev.*, **5**, 041108 (2018).

35. W. Ji, K. Yao, Y. C. Liang, *Adv. Mater.* **22**, 1763-1766 (2010).

36. O. Paull, C. Xu, X. Cheng, Y. Zhang, B. Xu, K. P. Kelley, A. de Marco, R. K. Vasudevan, L. Bellaiche, V. Nagarajan, D. Sando, *Nat. Mater.* **21**, 74-80 (2022).

37. J. Seidel, L. W. Martin, Q. He, Q. Zhan, Y.-H. Chu, A. Rother, M. E. Hawkridge, P. Maksymovych, P. Yu, M. Gajek, N. Balke, S. V. Kalinin, S. Gemming, F. Wang, G. Catalan, J. F. Scott, N. A. Spaldin, J. Orenstein, R. Ramesh, *Nat. Mater.* **8**, 229 (2009).

38. S. Farokhipoor, B. Noheda, *Phys. Rev. Lett.* **107**, 127601 (2011).

39. L. Qiao, S. Zhang, H. Y. Xiao, D. J. Singh, K. H. L. Zhang, Z. J. Liu, X. T. Zu, S. Li, *J. Mater. Chem. C* **6**, 1239-1247 (2018).

40. S. Manipatruni, D. E. Nikonov, I. A. Young, *Nat. Phys.* **14**, 338 (2018).

41. N. Balke, B. Winchester, W. Ren, Y. H. Chu, A. N. Morozovska, E. A. Eliseev, M. Huijben, R. K. Vasudevan, P. Maksymovych, J. Britson, S. Jesse, I. Kornev, R. Ramesh, L. Bellaiche, L.-Q. Chen, S. V. Kalinin, *Nat. Phys.* **8**, 81-88 (2012).

42. Y. Li, Y. Jin, X. Lu, J.-C. Yang, Y.-H. Chu, F. Huang, J. Zhu, S.-W. Cheong, *npj Quantum Mater.* **2**, 43 (2017).

43. J. Wang, J. Ma, H. Huang, J. Ma, H. M. Jafri, Y. Fan, H. Yang, Y. Wang, M. Chen, D. Liu, J. Zhang, Y.-H. Lin, L.-Q. Chen, D. Yi, C.-W. Nan, *Nat. Commun.* **13**, 3255 (2022).

44. J. Ma, J. Ma, Q. Zhang, R. Peng, J. Wang, C. Liu, M. Wang, N. Li, M. Chen, X. Cheng, P. Gao, L. Gu, L.-Q. Chen, P. Yu, J. Zhang, C.-W. Nan, *Nat. Nanotechnol.* **13**, 947-952 (2018).

45. M. A. P. Gonçalves, C. Escorihuela-Sayalero, P. Garca-Fernández, J. Junquera, J. Íñiguez, *Sci. Adv.* **5**, 2 (2019).

46. G. F. Nataf, M. Guennou, J. M. Gregg, D. Meier, J. Hlinka, E. K. H. Salje, J. Kreisel, *Nat. Rev. Phys.* **2**, 634-648 (2020).

47. G. Liu, Q. Zhang, H. H. Huang, P. Munroe, V. Nagarajan, H. Simons, Z. Hong, L.-Q. Chen, *Adv. Mater. Interfaces.* **3**, 1600444 (2016).

48. D. Zhang, D. Sando, P. Sharma, X. Cheng, F. Ji, V. Govinden, M. Weyland, V. Nagarajan, J. Seidel, *Nat. Commun.* **11**, 349 (2020).

49. S. L. Hsu, M. R. McCarter, C. Dai, Z. Hong, L.-Q. Chen, C. T. Nelson, L. W. Martin, R. Ramesh, *Adv. Mater.* **31**, 1901014 (2019).

50. H. N. Lee, H. M. Christen, M. F. Chisholm, C. M. Rouleau, D. H. Lowndes, *Nature* **433**, 395-399 (2005).

51. C. Lichtensteiger, *J. Appl. Crystallogr.* **51** (Pt 6), 1745 (2018).

52. P. Milde, D. Köhler, J. Seidel, L. M. Eng, A. Bauer, A. Chacon, J. Kindervater, S. Mühlbauer, C. Pfleiderer, S. Buhrandt, C. Schütte, A. Rosch, *Science* **340**, 1076-1080 (2013).

53. Q. Zhang, S. Prokhorenko, Y. Nahas, L. Xie, L. Bellaiche, A. Gruverman, N. Valanoor, *Adv. Funct. Mater.* **29**, 1808573 (2019).

54. A. Y. Abid, Y. Sun, X. Hou, C. Tan, X. Zhong, R. Zhu, H. Chen, K. Qu,. Y. Li, M. Wu, J. Zhang, J. Wang, K. Liu, X. Bai, D. Yu, X. Ouyang, J. Wang, J. Li & P. Gao *Nat. Commun.* **12**, 2054 (2021).

55. M. D. Rossell, R. Erni, M. P. Prange, J.-C. Idrobo, W. Luo, R. J. Zeches, S. T. Pantelides, R. Ramesh, *Phys. Rev. Lett.* **108**, 047601 (2012).

56. S. R. Bakaul, S. Prokhorenko, Q. Zhang, Y. Nahas, Y. Hu, A. Petford-Long, L. Bellaiche, N. Valanoor, *Adv. Mater.* **33**, 2105432 (2021).

57. S. Prosandeev, D. Wang, W. Ren, J. Íñiguez, L. Bellaiche, *Adv. Funct. Mater.* **23**, 234-240 (2013).

58. D. Albrecht, S. Lisenkov, W. Ren, D. Rahmedov, I. A. Kornev, L. Bellaiche, *Phys. Rev. B* **81**, 140401 (2010).

59. F. Borodavka, J. Pokorny, J. Hlinka, *Phase Transit.* **89**, 746-751 (2016).

60. O. Diéguez, P. Aguado-Puente, J. Junquera, J. Íñiguez, *Phys. Rev. B*, **87**, 024102 (2013).

61. J. Hlinka, M. Paściak, S. Körbel, P. Marton, *Phys. Rev. Lett.* **119**, 057604 (2017).

62. S. R. Burns, O. Paull, J. Juraszek, V. Nagarajan, D. Sando, *Adv. Mater.* **32**, 2003711 (2020).

63. D. Rahmedov, D. Wang, J. Íñiguez, L. Bellaiche, *Phys. Rev. Lett.* **109**, 037207 (2012).

64. S. Farokhipoor, C. Magén, S. Venkatesan, J. Íñiguez, C. J. Daumont, D. Rubi, E. Snoeck, M. Mostovoy, C. de Graaf, A. Müller, M. Döblinger, C. Scheu, B. Noheda, *Nature* **515**, 379-383 (2014).

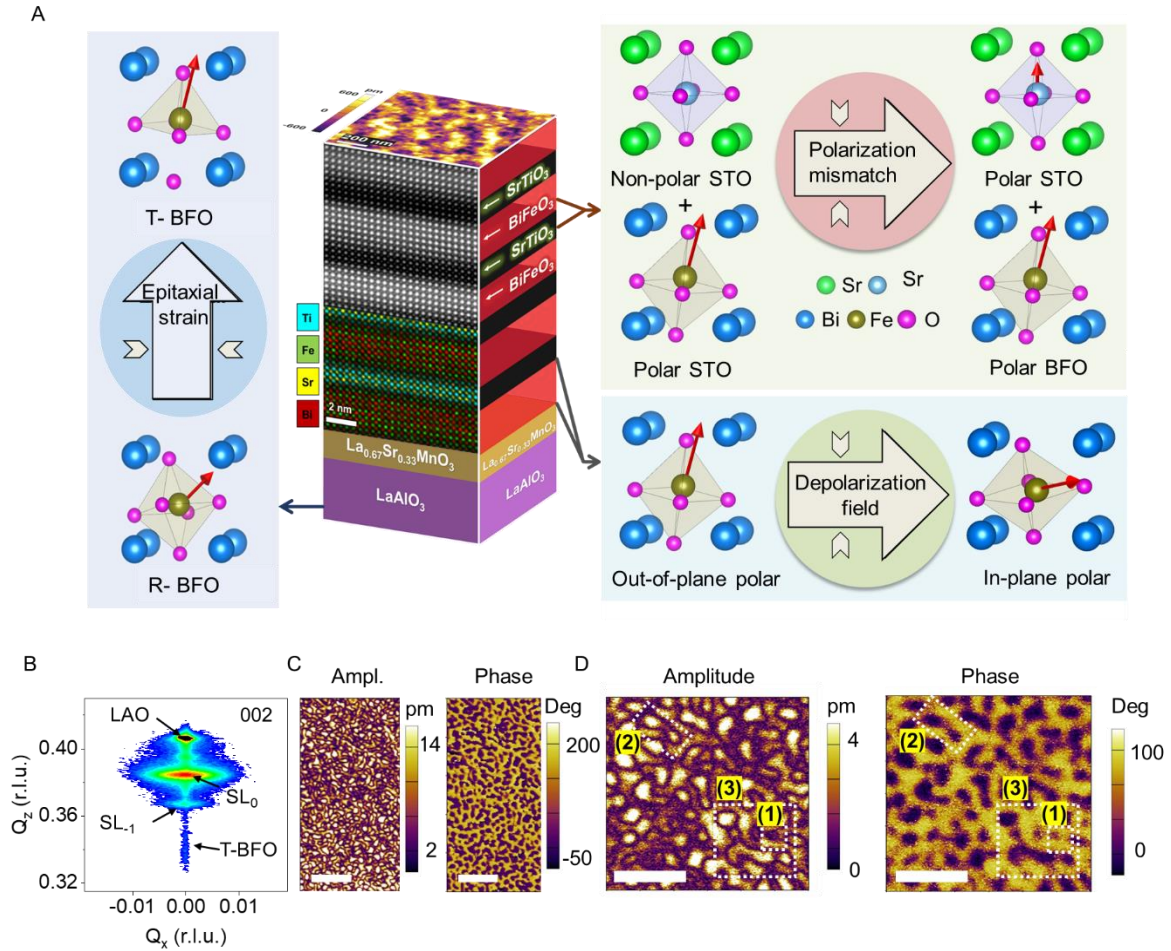

**Figure 1. Design of BFO/STO superlattice system and the observation of solitons in (BFO$_7$/STO$_4$)$_{10}$.** (**A**) Schematics of the superlattice with superimposed atomic resolution cross-sectional HAADF-STEM image and atomic EDS-mapping of Ti, Fe, Sr and Bi showing sharp interfaces. (left) LAO substrate induces a compressive epitaxial strain favoring a tetragonal phase. (right) The introduction of STO has two effects: (1) a polarization mismatch is introduced at BFO/STO interface causing STO to become polar. (2) the polarization discontinuity enhances the depolarization field causing polarization to curl. (**B**) Symmetrical x-ray diffraction reciprocal space map near the (002) reflection, showing superlattice peaks (**C**) PFM Amplitude and PFM phase images. Scale bars: 200 nm (**D**) Zoomed in PFM amplitude and phase images depicting topological structures such as (1) skyrmion (2) bimeron and (3) disclination. Scale bars: 100 nm.

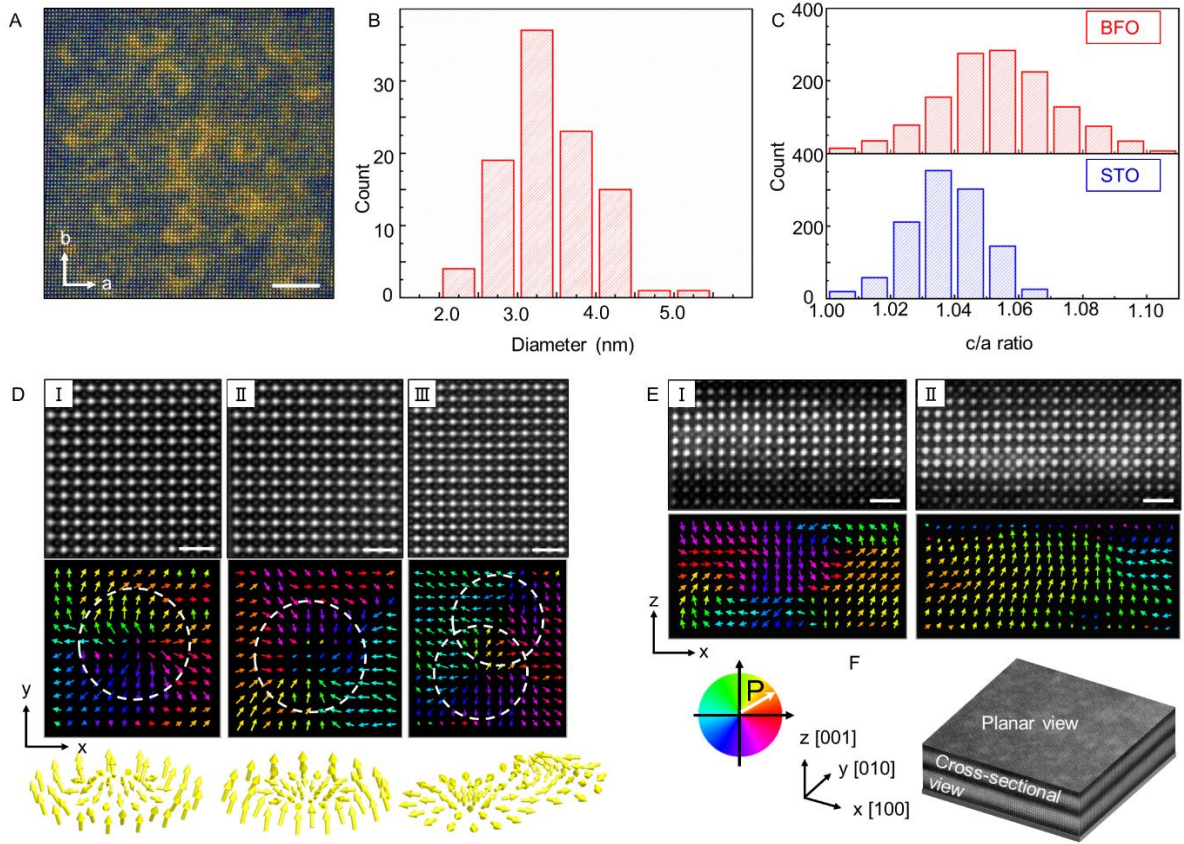

**Figure 2. Observation of ferroelectric solitons in BFO-STO superlattices.** (**A**) Low-magnification planar-view HAADF-STEM image of the $(BFO_7/STO_4)_{10}$ superlattice. Scale bar, 5 nm. (**B**) Histogram of size distribution of circular features. (**C**) The statistics of c/a ratios of STO and BFO extracted from the cross-sectional HAADF-STEM images. (**D**) Enlarged planar-view STEM-HAADF image and corresponding polar vectors, showing center-divergent, center-convergent and bimeron polar textures. Scale bar: 1 nm. (**E**) Enlarged cross-sectional STEM-HAADF images and corresponding polar vectors, showing a domain with anti-parallel (up–down) polarization and a trapezoidal domain with the convergent polar configuration. Scale bar: 1 nm. (**F**) Schematics of a planar-view HAADF-STEM image (30 nm × 30 nm) overlaid with a cross-sectional HAADF-STEM image, displaying an overview of the $(BFO_7/STO_4)_{10}$ superlattice.

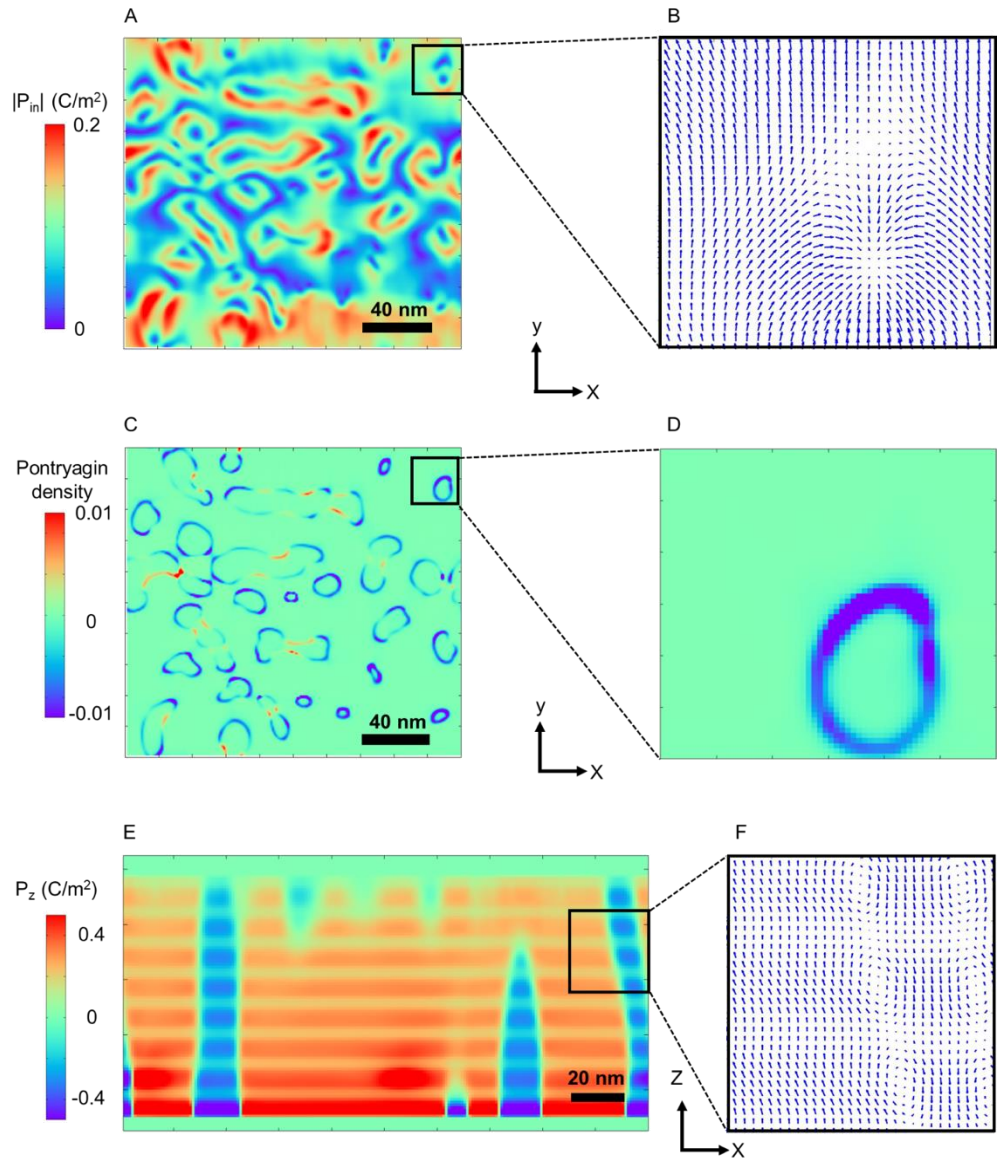

**Figure 3. Polar structures and Pontryagin density of (BFO$_7$/STO$_4$)$_8$ superlattices as calculated from phase-field simulations.** (**A**) Planar view of the in-plane polarization magnitude. (**B**) Magnified view of the in-plane polar vector plot, showing a bimeron structure. (**C**) Corresponding plot of the Pontryagin density. (**D**) Magnified view of the Pontryagin density of the bimeron structure. Surface integration of the Pontryagin density for the bimeron gives a topological charge of -1. (**E**) Cross-section view of the out-of-plane polarization, showing alternating positive and negative polarization. (**F**) Magnified view of the out-of-plane polar vector plot, showing a vortex-like structure, with a mixture of *R*-like and *T*-like polar regions.

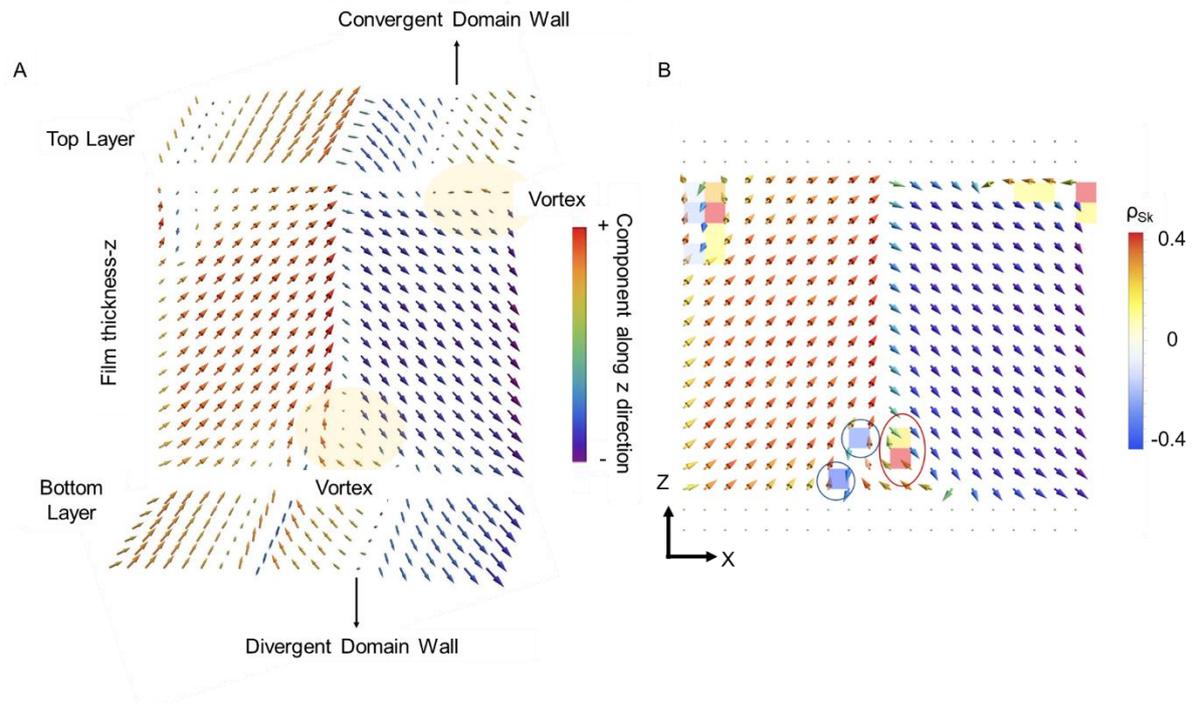

**Figure 4. (A)** Polar structure (arrows) of BiFeO$_3$ thin film supercell after relaxation from a 109° domain structure under -2.9% epitaxial strain at 10 K. The x-z plane is shown along with the top and bottom most layers of the film. **(B)** The distribution of the Pontryagin's charge density (colored plaquettes) along with the normalized dipoles (arrows) for the polar distribution shown in A.